\documentclass[aps, twocolumn, showpacs, amsmath,superscriptaddress]{revtex4}
\usepackage{dcolumn}
\usepackage{bm}
\usepackage{graphicx}
\usepackage{color}
\usepackage{ulem}

\begin{document}
\def\be{\begin{equation}}
\def\ee{\end{equation}}

\def\bc{\begin{center}} 
\def\ec{\end{center}}
\def\bea{\begin{eqnarray}}
\def\eea{\end{eqnarray}}
\newcommand{\avg}[1]{\langle{#1}\rangle}
\newcommand{\ket}[1]{\left |{#1}\right \rangle}
\newcommand{\beq}{\begin{equation}}
\newcommand{\eneq}{\end{equation}}
\newcommand{\beqnn}{\begin{equation*}}
\newcommand{\eneqnn}{\end{equation*}}
\newcommand{\beqy}{\begin{eqnarray}}
\newcommand{\eneqy}{\end{eqnarray}}
\newcommand{\beqynn}{\begin{eqnarray*}}
\newcommand{\eneqynn}{\end{eqnarray*}}
\newcommand{\half}{\mbox{$\textstyle \frac{1}{2}$}}
\newcommand{\proj}[1]{\ket{#1}\bra{#1}}
\newcommand{\av}[1]{\langle #1\rangle}
\newcommand{\braket}[2]{\langle #1 | #2\rangle}
\newcommand{\bra}[1]{\langle #1 | }

\newcommand{\inprod}[2]{\braket{#1}{#2}}
\newcommand{\upket}{\ket{\uparrow}}
\newcommand{\downket}{\ket{\downarrow}}
\newcommand{\Tr}{\mathrm{Tr}}
\newcommand{\hcontrol}{*!<0em,.025em>-=-{\Diamond}}
\newcommand{\hctrl}[1]{\hcontrol \qwx[#1] \qw}
\newenvironment{proof}[1][Proof]{\noindent\textbf{#1.} }{\ \rule{0.5em}{0.5em}}
\newtheorem{mytheorem}{Theorem}
\newtheorem{mylemma}{Lemma}
\newtheorem{mycorollary}{Corollary}
\newtheorem{myproposition}{Proposition}
\newcommand{\vp}{\vec{p}}
\newcommand{\Or}{\mathcal{O}}
\newcommand{\so}[1]{{\ignore{#1}}}

\newcommand{ \red}[1]{\textcolor{red}{#1}}
\newcommand{\blue}[1]{\textcolor{blue}{#1}}

\title{Phase transition of light on complex quantum networks}

\author{Arda Halu}

\affiliation{Department of Physics, Northeastern University, Boston, 
Massachusetts 02115 USA}

\author{Silvano Garnerone}

\affiliation{Institute for Quantum Computing, University of Waterloo, Waterloo, ON N2L 3G1, Canada}

\author{Alessandro Vezzani}

\affiliation{Dipartimento di Fisica, Universit\'a degli Studi di Parma, V.le G.P. Usberti n.7/A, 43100 Parma, Italy}
\affiliation{
Centro S3, CNR Istituto di Nanoscienze, via Campi 213/a, 41100 Modena, Italy}

\author{Ginestra Bianconi}

\affiliation{School of Mathematical Sciences, Queen Mary University of London, London E1 4NS, United Kingdom}
\begin{abstract}
Recent advances in {quantum} 
optics and atomic physics allow for an {unprecedented} 
level of control {over} light-matter 
interactions, {which can be exploited to investigate new physical phenomena}.
 {In this work we are interested in the role played by the topology of quantum networks describing coupled 
optical cavities and local atomic degrees of freedom}.
In particular{, using a mean-field approximation,} we study {the phase diagram of} the Jaynes-Cummings{-Hubbard} model on complex networks topologies, and we characterize 
 the transition between a Mott-like phase of localized polaritons and a superfluid phase.
We found that, { for complex topologies,} the phase diagram is non-trivial and well defined in the thermodynamic limit only if the hopping {coefficient} scales like {the} inverse of the maximal eigenvalue of the {adjacency} matrix of the network.
 {Furthermore we provide numerical} evidences that, for {some} complex network topologies, this {scaling} implies {an asymptotically vanishing hopping coefficient} in the limit of large network sizes. {The latter result suggests the interesting possibility of observing quantum phase transitions of light on complex quantum networks even with very small couplings between the optical cavities}.
\end{abstract}


\maketitle
\section{Introduction}
Quantum optics and atomic physics have reached a level of 
control over light-matter interactions which {not only
makes it} feasible the emulation of  
condensed matter models \cite{Plenio,Plenio2}, {but also inspires} new 
architectures envisaging a 
future quantum Internet with a desired topology \cite{QI, Ritter}. 
 {Hence} the potential advantage 
 {coming from a combined optical and atomic approach is twofold:} on one hand, 
being able to control a quantum system that 
simulates another one is a way to realize a special purpose 
quantum computer; on the other hand, the 
possibility  {of manipulating} new degrees of freedom (not accessible 
in condensed matter systems) 
motivates the experimental and theoretical study of new quantum 
systems, {with the possibility of discovering new physical phenomena}. 
 {In this respect an important outcome, coming from the combined experimental investigations 
of atomic and optical systems, is the realization of}
coupled cavity arrays
interacting with {local} atomic degrees of freedom \cite{Plenio}. 
 {These systems allow for the controlled interaction between trapped atoms and local cavity photons, 
and moreover photons are free to hop 
between coupled cavities.} 
 {Changing the details of the physical setup different many-body models can be realized} \cite{Plenio,Schiro},
 {and one in particular is of interest in the present work:}
the Jaynes-Cummings-Hubbard (JCH) 
model \cite{Jaynes, Greentree, Greentree2, LeHur,Vuckovic,Blatter,Pelster}. 
Part of the interest in these systems is motivated by the 
possibility to investigate {new} quantum critical {phenomena} \cite{Sachdev}, like the quantum phase transition of light between 
a Mott-like regime 
and a superfluid phase \cite{Jaynes, Greentree, Greentree2,LeHur, Vuckovic,Blatter, Pelster}. Also of interest is the possibility to generate quantum simulators that naturally access non-equilibrium physics \cite{Koch}.

In this work we are interested in an additional degree of freedom which optical 
arrays can provide, i.e. the topology of the network underlying the quantum dynamics. While regular lattice structures with short-range interactions are the typical framework in standard quantum emulator architectures, fiber-coupled cavities may allow for the 
realization of quantum networks with distant effective interactions between local degrees of freedom \cite{Ritter,Zheng, Kyoseva}. These effective long-range interactions are an important ingredient for the construction of quantum networks with complex 
topology, which is a recent topic of interest in the quantum information and complex network community 
\cite{Ranking_Silvano1, Ranking_Silvano2, Ranking_Jesus, Jesus_Entaglement, 
Burioni_BEC,Havlin_localization_1, 
Havlin_localization_2, Hubbard_Apollonian,Free_Electron_Gas_Apollonian,Bose_Einstein_Apollonian,QTIM, 
JSTAT,BH,Severini,entropy_silvano}.
Along this line of research, in this work 
we study the effect of the array topology on the phase diagram of the JCH model. 
 {Motivations come not only from the possible experimental realization of such systems, but also from 
a number of results, {especially} 
in the classical context, underlying the importance of networks' topologies. Indeed}
for classical systems it is well known that the topology of the network can significantly change the phase diagram of  {some} models, and {their} critical behavior \cite{crit,Dynamics}. 
On the quantum side 
 {previous} results on the Bose-Hubbard model \cite{Fisher,Greiner, Inguscio,Vezzani,Laser} are 
particularly {inspiring} for the present work.
 {In fact, for this model,} 
it has been shown by mean-field arguments that the phase diagram depends of the 
maximal eigenvalue {$\Lambda$} of the hopping matrix describing the topology of the network \cite{BH,Laser}. 
This result is valid both in presence of disorder \cite{Laser} and in presence of a complex topology \cite{BH}. 
Interestingly in a complex random network topology, or in an Apollonian network \cite{Apollonian1,Apollonian2}, the 
maximal eigenvalue of the adjacency matrix diverges with the network size. 
 {In \cite{BH} it has been shown, for the Bose-Hubbard model, that this divergence implies a non-trivial scaling of the hopping coefficient, in order to not suppress the Mott-insulator phase in the thermodynamic limit}.

 {Here} we consider instead the properties of the {JCH} model on 
complex quantum network topologies.
 {Using mean-field theory we} characterize the phase diagram of the model at $T=0$, 
 which {presents a} phase transition between a Mott-like 
regime and a superfluid phase.
 {We demonstrate analytically, and confirm numerically, that the} 
phase diagram is non-trivial and well defined only if the hopping coefficient $\kappa$ scales as the inverse of the 
maximal eigenvalue $\Lambda$ of the hopping matrix, i.e. $\kappa\propto \frac{1}{\Lambda}$.
 {Furthermore} we characterize the scaling of {the} maximal eigenvalue for a number of {well known} 
complex network topologies, showing that in many cases the maximal eigenvalue $\Lambda$ diverges with the 
network size $N$.
Therefore {our results are of general interest for a number of different complex topologies,} 
 {and they imply that for complex network arrays interesting quantum critical behaviors can be observed 
even with very small couplings between different cavities.}

 {The rest of the paper is organized as follows: in Sec.~\ref{sec:JC} we introduce the Jaynes-Cummings Hamiltonian and some of its properties; in Sec.~\ref{sec:JCH} we characterize the Jaynes-Cummings-Hubbard model, its mean-field solution, and we consider the scaling behavior of the hopping coefficient for different network topologies; Sec.~\ref{sec:conclusion} is devoted to discussions and conclusions. In the Appendix a detailed derivation of the mean field solution is provided.}

\section{Atom-Photon interaction in a single cavity }
\label{sec:JC}
\begin{figure}
\center
\includegraphics[width=1.0\columnwidth ]{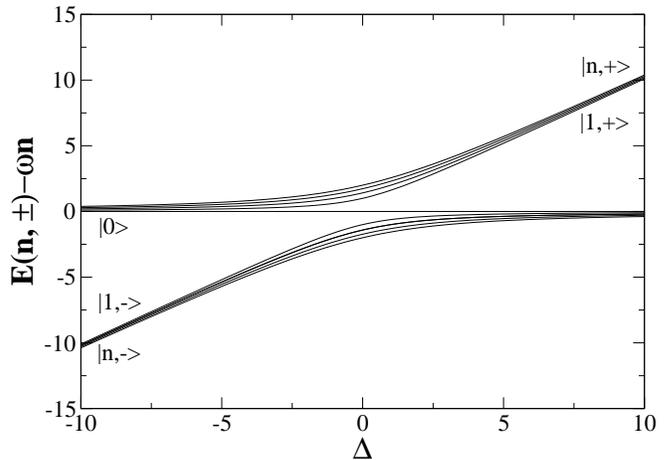}
\caption{The energy non-linear dependence of the spectrum $E(n,\pm)-n\omega$ as a function of the detuning $\Delta$ for $\beta=1$. The non-linearity effects are stronger for low values of $n$.}
\label{photonB.fig}
\end{figure}

 {The standard model describing the interaction between a two-level atom and quantized 
electromagnetic modes is provided by the Jaynes-Cummings Hamiltonian.}
In the rotating wave approximation, {and assuming a single cavity mode}, 
the Hamiltonian 
is given by
\begin{equation}
H^{JC}=\epsilon \sigma^{+}\sigma^{-}+\omega a^{\dag}a+\beta (\sigma^+a+\sigma^- a^{\dag})
\end{equation}
where $\epsilon$ is the atomic transition frequency, $\omega$ is the field frequency, and 
$\beta$ is the atom-cavity coupling constant; $a$ and $a^{\dag}$ are the bosonic lowering and 
raising operators, while $\sigma^{\pm}$ are the atomic lowering and raising operators of the 
two level system.
The eigenstate of this Hamiltonian are {\it polaritons}, or {\it dressed states}, 
 {given by} a combination of atom and field states. In the base $\left\{\ket{0,\downarrow}, \ket{0,\uparrow},\ket{1,\downarrow},\ket{1,\uparrow}\ldots \right\}$, the atom states 
 {are represented in the basis of the eigenstates $\ket{\downarrow}, \ket{\uparrow}$ of the 
Pauli $\sigma_z$ operator, while the field states are denoted with the number operator's eigenstates $\ket{n}$.
The Jaynes-Cummings Hamiltonian eigenstates are given by} 
\bea
\ket{n,-}=\cos \theta_n \ket{n,\downarrow}-\sin \theta_n \ket{n-1,\uparrow} \nonumber \\
\ket{n,+}=\sin \theta_n \ket{n,\downarrow}+\cos \theta_n \ket{n-1,\uparrow}
\label{eigve}
\eea
for every $n\geq1$, where the angle $\theta_n$ is expressed in terms of the detuning parameter 
 $\Delta=\epsilon-\omega$ and is given by 
\bea
\theta_n=\frac{1}{2}\arctan \left(\frac{2\beta\sqrt{n}}{\Delta}\right).
\eea
 {The eigenvalues associated to these eigenstates are given by}
\bea
E(n,\pm)=\omega n+\frac{\Delta}{2}\pm \sqrt{n\beta^2+\frac{\Delta^2}{4}}
\label{eigva}
\eea
 {In addition to the above dressed state, another eigenstate of the system is  {$\ket{0}\equiv\ket{0,\downarrow}$},
when $n=0$, with the associated eigenvalue {$E_0=0$}}.
The fundamental state of the system should be determined for every fixed value of the parameters.
If we proceed in this calculation we can observe first of all that the ground state will be either the state with zero polations $\ket{n=0}$ and energy $E_{0}=0$, or one of the states $\ket{n-}$  {associated to the} eigenvalues 
$E_{n,-}$. Indeed, for every fixed number of polaritons $n\geq 1$ we have $E_{n,+}>E_{n,-}$.
If we consider the spectrum in the limit $\omega\gg|\Delta|,\beta$ the state with zero 
polaritons $\ket{n=0}$ will be the ground state. As we decrease $\omega$ we will find a point 
 {in the parameter space}
where $E_0=E_{1,-}$, precisely at $\omega=\Delta/2-\sqrt{(\Delta/2)^2+\beta^2}$. Lowering the value of 
$\omega$ further we {will} find a full set of degeneracy points given by (for $n\geq 1$)
\bea
\frac{\omega}{\beta}=\sqrt{n+1+\left(\frac{\Delta}{2\beta}\right)^2}-\sqrt{n+\left(\frac{\Delta}{2\beta}\right)^2}.
\eea
The energy spectrum of the atom-cavity system, given by Eq. $(\ref{eigva})$, has a nonlinear dependence on $n$ (see the energy spectrum in Figure $\ref{photonB.fig}$). This anharmonicity
in the splitting of the energy eigenstates gives rise to nonlinear phenomena at the single-photon level. One of the most relevant {of these} is {\it photon blockade}, where the presence of one photon stops further absorption of photons from a coherent light source \cite{Deutsch,Kimble,Bose}.

\section{Jaynes-Cummings-Hubbard model}\label{sec:JCH}
 {Optical cavities, with trapped atoms, can be arranged in arrays where the overlap between different cavities 
wave-functions allow photons to hop from one site to another. The Hamiltonian describing this new physical scenario 
is now known as Jaynes-Cummings-Hubbard model.
The inclusion of optical fibres, or other optical devices, 
can be used to realize more complex geometries, where the hopping is not restricted to nearest neighbour cavities on 
a regular lattice \cite{Ritter,Zheng, Kyoseva}. This is precisely the kind of situation that we want to investigate in this paper.}

To tune the number of polaritons in each cavity a chemical potential $\mu$ 
 {might be used}, hence
 the full {JCH} model 
 {will be described by} the following Hamiltonian, 
\begin{equation}
H^{JCH}=\sum_i \left[H_i^{JC}-\mu N_i \right]+H^{hop}
\label{eq:JCH}
\end{equation}
where 
\bea
H^{JC}_i=\epsilon \sigma^{+}_i \sigma^{-}_i+\omega a^{\dag}_ia _i+\beta (\sigma^+_i a_i+\sigma^-_i a^{\dag}_i)
\eea
is the Jaynes-Cummings Hamiltonian for a single cavity, and
$N_i$ indicates the number of polaritons in each cavity 
($
N_i=\sigma^+_i\sigma^-_i+a^{\dag}_ia_i
$); 
while $H^{hop}$ is the hopping term.
 {Note} that the chemical potential $\mu$ is not an experimentally tunable parameter for this system. In real experiments appropriate preparation schemes have to be devised in order to obtain states with different polariton number. Recently,  {using the Rabi model}, it has been shown that the inclusion of counter rotating terms can stabilize finite-density quantum phases of correlated photons without the use of a chemical potential \cite{Schiro}. 
 {The last hopping term in Eq. (\ref{eq:JCH})} is characterized by a the strength $\kappa$ and the adjacency matrix of the underlying quantum network ${\tau}$, and is given by 
\bea
H^{hop}=-\kappa \frac{1}{2}\sum_{i,j}\tau_{ij} (a^{\dag}_ia_j+a^{\dag}_ja_i).
\eea

 {Let us consider first two extreme cases: one in which the hopping strength is very small, and the other where the atom-photon interaction is negligible.}
In the atomic limit $\kappa/\beta\ll 1$, the Hamiltonian $H^{JCH}$ becomes, to first order 
approximation, the sum of single cavity Hamiltonians $H=\sum_i H_i$ {,} with $H_i$ given by 
\bea
H_i=H_i^{JC}-\mu N_i.
\eea
The eigenstates of the single cavity Hamiltonian are given by Eqs. $(\ref{eigve})$ for $n\geq 1$ {,} 
and $\ket{0,\downarrow}=\ket{0}$ for $n=0$.
The corresponding eigenvalues are 
 \bea
 E_{n\pm}^{\mu}=(\omega-\mu) n+\frac{\Delta}{2}\pm \sqrt{n\beta^2+\frac{\Delta^2}{4}},
\label{Emu}
\eea
for $n\geq1$, and $E_{0}=0$ for $n=0$.
The ground state of the system can be calculated similarly to the case of a single cavity. Indeed, for every cavity we will found that the ground state is constituted either by the eigenstate $\ket{n, -}$ or by the eigenstate $\ket{0}$.

In the hopping dominated limit $\kappa/\beta \gg1$ we 
 {can treat perturbatively} the atom-photon interaction.
 {$H^{JCH}$ reduces, to first-order approximation, to} 
a tight-binding hamiltonian $H^{tb}$ given by 
\bea
H^{tb}=\sum_i (\omega-\mu) a_i^{\dag}a_i-\kappa \frac{1}{2}\sum_{i,j}\tau_{ij} (a^{\dag}_ia_j+a^{\dag}_ja_i) {.}
\eea
 {The eigenvalues of $H^{tb}$ depends in a simple way from the eigenvalues $\lambda_n$ of the 
adjacency matrix of the quantum network:}

\bea
E_n=N(\omega-\mu -\kappa \lambda_n).
\eea
 {The above equation reveals an instability of the system for}
\bea
\kappa\Lambda>\omega-\mu
\eea
where $\Lambda$ is the maximal eigenvalue of the adjacency matrix ${\tau}$. 
From this result we can already conclude that the maximal eigenvalue of the adjacency matrix set an important scale for the strength of the hopping coefficient $\kappa$.

\subsection{Mean-field theory }
In order to explore the phase diagram of the Jaynes-Cummings-Hubbard model 
 {we make use of the mean-field treatment of the hopping term, which reduces to the following approximation}
\begin{equation}
a_ia^{\dag}_j\simeq\avg{a_i}a^{\dag}_j+a_i\avg{a^{\dag}_j}-\avg{a_i}\avg{a^{\dag}_j}.
\end{equation}
Therefore the hopping term {becomes}
\bea
H_{hop}^{MF}=-\kappa \sum_{i,j}\tau_{ij} (a^{\dag}_i+a_i)\psi_j+\kappa \sum_{i,j}\tau_{ij}\psi_i\psi_j,
\eea
where we have indicated by $\psi_i$ the local order parameter  $\psi_i \equiv\avg{a_i}$ 
 {(also equal to $\avg{a^{\dag}_i}$, due to the gauge symmetry of the model)}.
This Hamiltonian displays a phase transition between a Mott-Insulator phase, where $\psi_i=0$ $\forall i$, and a superfluid phase.
In order to study the phase diagram of this model, within the mean-field approximation we treat $H^{hop}$ as a perturbation and we calculate $\psi_i$ self-consistently, to first order in $\kappa$,
obtaining (see the Appendix for more details)
\bea
\psi_i=\kappa \sum_j \tau_{ij} \psi_j R_n,
\label{psii}
\eea
with $R_n$ given by 
\bea
R_0&=&\left[\frac{\cos^2 \theta_1}{E_{1-}^{\mu}}+\frac{\sin^2 \theta_1}{E_{1+}^{\mu}}\right]\\
R_{n \geq 1}&=&-\left[\frac{|\sqrt{n+1}\cos \theta_n \cos\theta_{n+1}+\sqrt{n}\sin \theta_n \sin \theta_{n+1}|^2}{E^{\mu}_{n-}-E^{\mu}_{(n+1)-}}\right.\nonumber \\
&+&\frac{|\sqrt{n+1}\cos \theta_n \sin\theta_{n+1}-\sqrt{n}\sin \theta_n \cos \theta_{n+1}|^2}{E^{\mu}_{n-}-E^{\mu}_{(n+1)+}}\nonumber \\
&+&\frac{|\sqrt{n}\cos \theta_n \cos\theta_{n-1}+\sqrt{n-1}\sin \theta_n \sin \theta_{n-1}|^2}{E^{\mu}_{n-}-E^{\mu}_{(n-1)-}}\nonumber \\
&+&\left.\frac{|\sqrt{n}\cos \theta_n \sin\theta_{n-1}-\sqrt{n-1}\sin \theta_n \cos \theta_{n-1}|^2}{E^{\mu}_{n-}-E^{\mu}_{(n-1)+}}\right]\nonumber
\eea
 {where the integer $n\geq 0$ depends on the systems parameters and it is chosen minimizing the on site energy $E_{n-}^{\mu}$given by Eq. (\ref{Emu}).}
If we diagonalize Eq.$(\ref{psii})$, along {with} the eigenvalues of the adjacency matrix $\tau$, we get that the critical line for the transition between the Mott-insulator phase and the superfluid phase is given by 
\bea
\kappa\Lambda R_n=1,
\label{crit}
\eea
where $\Lambda$ is the maximal eigenvalue of the adjacency matrix $\tau$.
 {This clearly shows that the phase diagram of the model depends}
on the product $\kappa\Lambda$.
On regular graphs we have $\psi_i=\psi$ $\forall i$, and $\Lambda=z$, {where $z$ is the connectivity of the lattice}.
 {On the other hand} for complex topology the maximal eigenvalue of the hopping matrix $\Lambda$ can be significantly different from the average connectivity of the networks.
In particular, for a large variety of networks the maximal eigenvalue of the adjacency matrix $\Lambda$ diverges with the network size.
This suggests that, in order to have a non-trivial phase diagram for the Jaynes-Cummings-Hubbard model, the hopping strength $\kappa$ must scale as
\bea
\kappa\propto \frac{1}{\Lambda}.
\label{scaling}
\eea 
In the following {section} we will investigate {for} different network topologies the {respective} scaling, with the network size, of the maximal eigenvalue {of the adjacency matrix} \cite{Chung, K}.
We note here that at the mean-filed level the phase boundary is given by Eq. ~$(\ref{crit})$.  
{With respect to the phase diagram on regular lattices, the effect of the complex topology 
 is the substitution of the average degree $z$ of the lattice with the maximal eigenvalue {$\Lambda$} of the network. Therefore the dependence of the phase boundary on the detuning
parameter $\Delta$ is similar to the one observed in regular lattices \cite{Pelster}. In fact, 
{as soon as the detuning is different from zero,} the Mott lobes with mean polariton number {greater} than one are reduced in size and shifted to smaller value of the chemical potential. {This effect is} independent on the sign of the detuning parameter. We remark that on complex network as in regular lattices the thermal
fluctuations destroy the Mott insulator phase. Therefore at finite
temperature the phase diagram should be composed by a superfluid regime
and a normal fluid. }

\begin{figure}
\centering
$\begin{array}{c }
{\includegraphics[width=0.8\columnwidth]{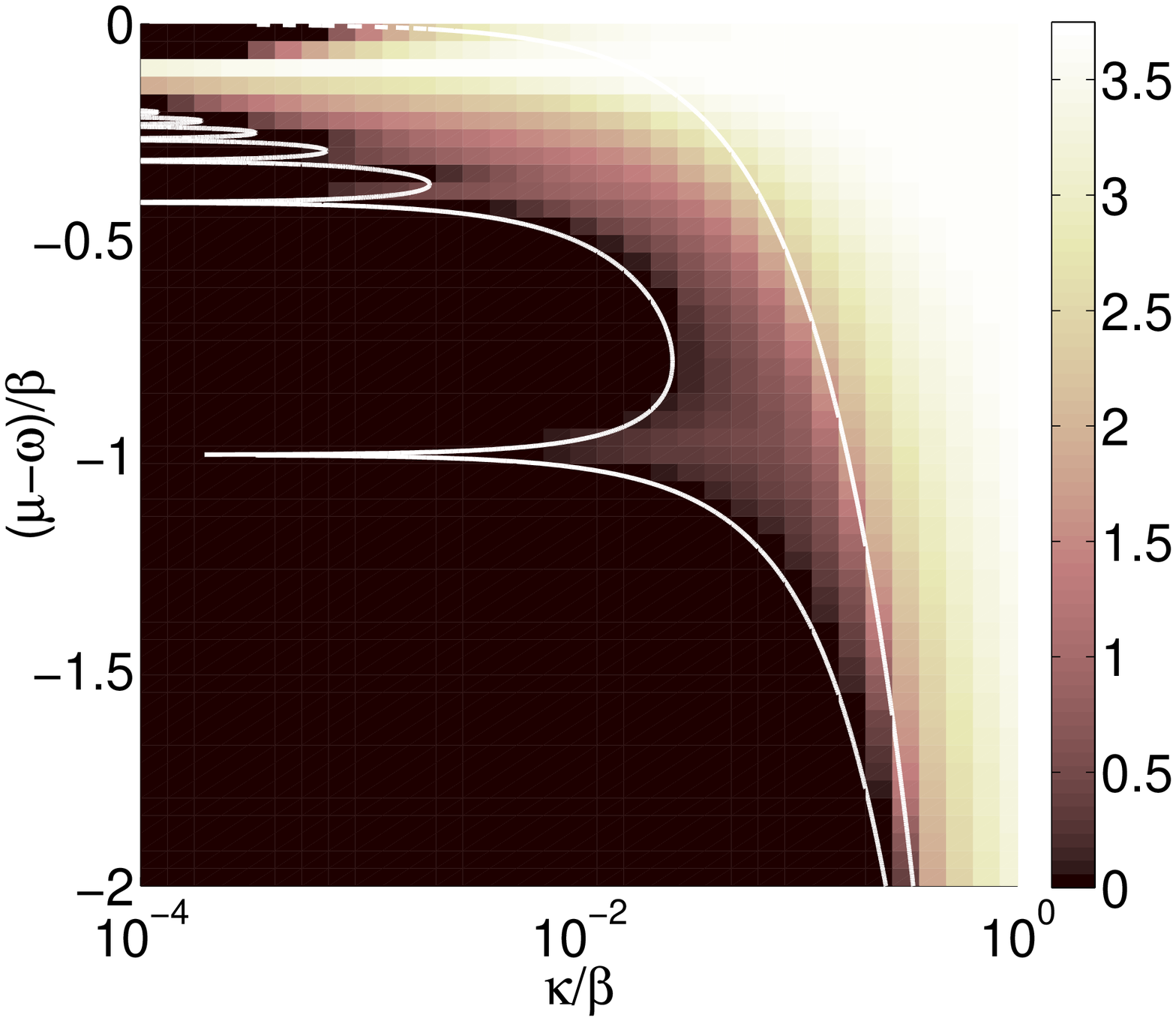}} \nonumber \\
{\includegraphics[width=0.8\columnwidth]{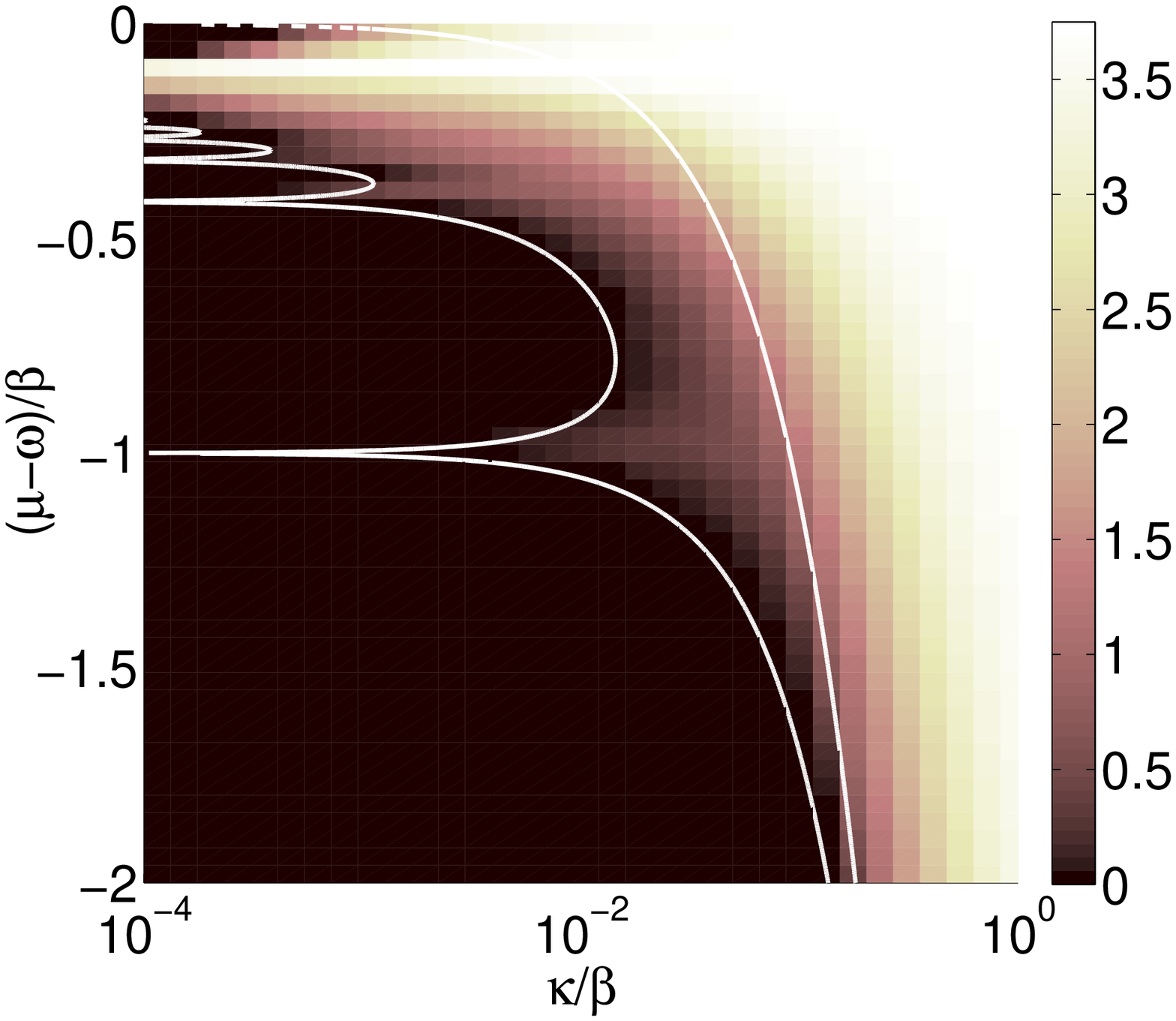}}\nonumber \\
{\includegraphics[width=0.8\columnwidth]{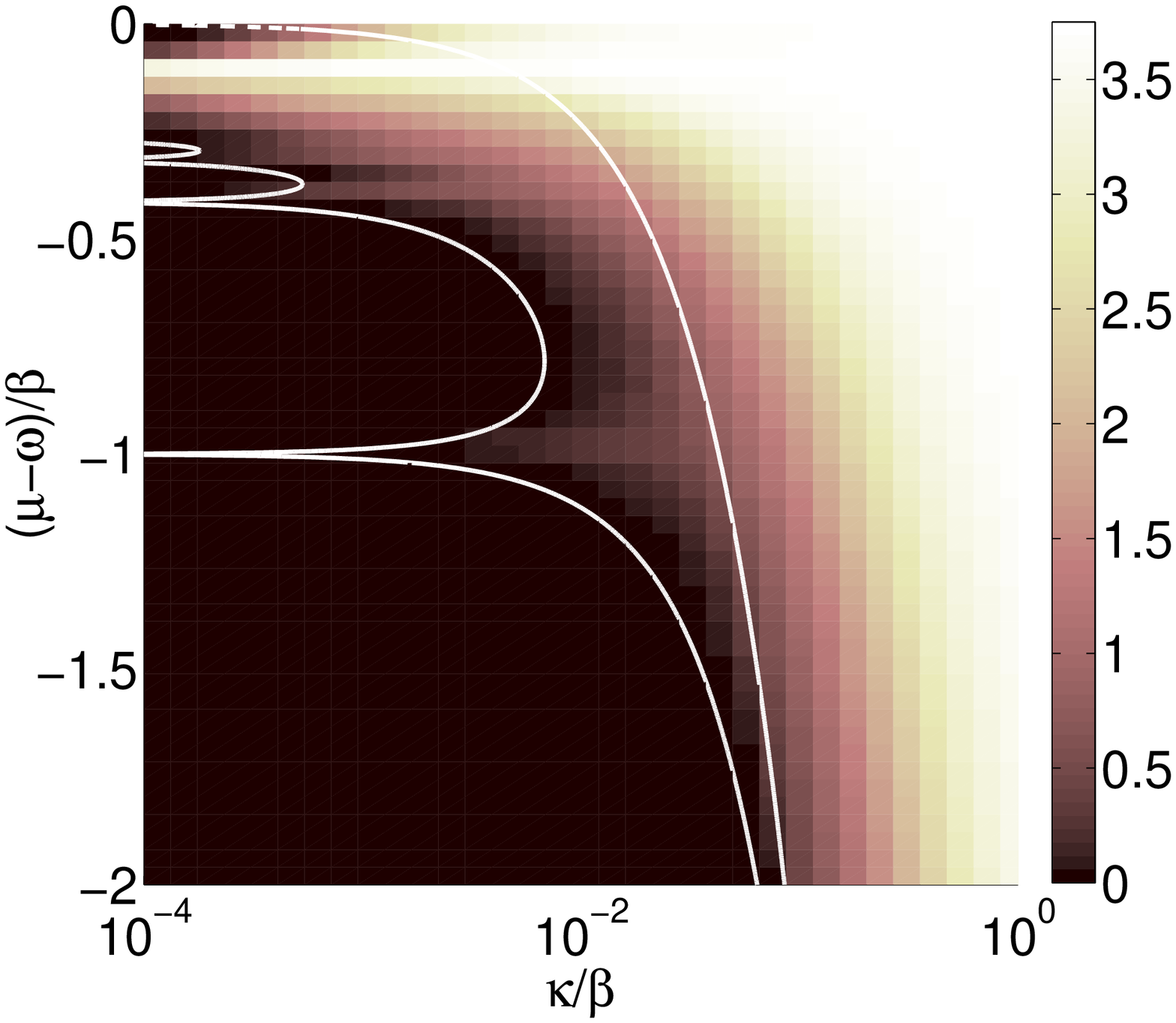}}\nonumber \\
\end{array}$
\caption{(Color online) Mean-field phase diagram of the Jaynes-Cummings-Hubbard model with $\Delta=0$ on a random scale-free network with power-law exponent $\gamma=2.2$ for different value of $N$, i.e. $N=100$ (top panel) $N=1000$ (middle panel) $N=10000$ (bottom panel). {The phase diagram scales with the maximal eigenvalue} that is given by $\Lambda=5.98$ (top panel) $\Lambda=11.07$ (middle panel) and by $\Lambda=23.42$ (bottom panel). {The solid lines denote the analytic perturbative solution in mean field of the model.}}
\label{fig:scalefree}
\end{figure}

\subsection{Regular networks}
For regular networks and regular lattices with connectivity $z$ the maximal eigenvalue of the adjacency matrix $\Lambda$ is independent on the network size. In particular, we have \bea
\Lambda=z.\eea
 In this {case} the critical line Eq. $(\ref{crit})$ coincides with the one
found in the literature using the mean-field approximation \cite{Greentree, Greentree2, LeHur}.

\subsection{Random graphs}
For random Erd\"os-Renyi graphs with finite connectivity and Poisson degree distribution, it has been proven \cite{K} that 
\bea
\Lambda\propto \sqrt{k_{max}},
\eea
where $k_{max}$ is the maximal degree of the system.
For random networks with a finite connectivity we have $k_{max}=\ln N/ \ln\ln N$, therefore 
\bea
\Lambda(N)\propto \sqrt{\frac{\ln N}{\ln \ln N}}. 
\label{ER}
\eea
Considering the scaling given by Eq.$(\ref{scaling})$, we have that the hopping strength has to satisfy the following relation in order to have a non-trivial phase diagram
\bea
\kappa(N) \propto \sqrt{\frac{\ln \ln N}{\ln N}}. 
\label{ER2}
\eea
\subsection{Random scale-free networks}
 {Scale-free networks provide one of the most interesting and most studied 
 topology for the {analysis} of phase 
transitions occurring on them.}
 {Indeed, classically, on scale-free networks with degree distribution $P(k)\propto k^{-\gamma}$ and power-law 
exponent $\gamma\in(2,3]$ the phase diagram of the Ising model \cite{ising1, ising2, ising3,Bradde}, and the 
percolation transition \cite{percolation1, percolation2} change drastically due to the diverging second moment of the 
average degree $\avg{k^2}$. A similar observation can be also made for the epidemic spreading model on annealed 
complex networks \cite{epidemics}, i.e. complex networks in which the links are dynamically rewired.
Moreover, the spectral properties of the complex networks determine the critical behavior of the epidemic spreading 
on complex quenched networks \cite{Durrett, epidemics2} {for} the $O(N)$ model \cite{Burioni_ON1,Burioni_ON2} and 
the stability of the synchronization dynamics \cite{Synchr1,Synchr2}.}

In random  scale-free networks, with power-law degree distribution $P(k)\propto k^{-\gamma}$ it has been proven \cite{Chung} that the maximal eigenvalue scales like
\bea
\Lambda\propto\left\{\begin{array}{ccc}\sqrt{k_{max}} &\mbox{for} &\gamma>2.5\nonumber \\
\frac{\avg{k^2}}{\avg{k}} & \mbox{for} &\gamma<2.5.
\end{array}\right.
\eea
Moreover, the maximal degree of the network satisfy $k_{max}=\min\left[{N^{1/2},N^{1/(\gamma-1)}}\right]$, where we have considered the structural cutoff of the degrees of the network for $\lambda\leq3$.
Therefore, the maximal eigenvalue of the network $\Lambda$ follow a different scaling with the network size, depending of the power-law exponent $\gamma$,
\bea
\Lambda(N)\propto\left\{\begin{array}{ccc} N^{1/[2(\gamma-1)]} &\mbox{for} &\gamma>3\nonumber \\
N^{1/4}&\mbox{for} &2.5<\gamma\leq3\nonumber \\
N^{(3-\gamma)/2} & \mbox{for} &\gamma<2.5.
\end{array}\right.
\eea 
Finally, the hopping coefficient $\kappa$ that ensure a non-trivial phase diagram [see Eq. $(\ref{scaling})$] scales with the network 
size $N$ according to the following rules
\bea
\kappa(N)\propto\left\{\begin{array}{ccc} N^{-1/[2(\gamma-1)]} &\mbox{for} &\gamma>3\nonumber \\
N^{-1/4}&\mbox{for} &2.5<\gamma\leq3\nonumber \\
N^{-(3-\gamma)/2} & \mbox{for} &\gamma<2.5
\end{array}\right.
\eea
 {Figure\ref{fig:scalefree} shows both the analytic perturbative solution in mean-field of the JCH model, and the numerical non-perturbative mean-field evaluation of the phase diagram. As can be seen there is substantial agreement between the two, and furthermore one can observe the dependence on the size of the network of the detailed location of the critical lines.} 

 \begin{figure}
\centering
$\begin{array}{c c}
{\includegraphics[width=0.45\columnwidth]{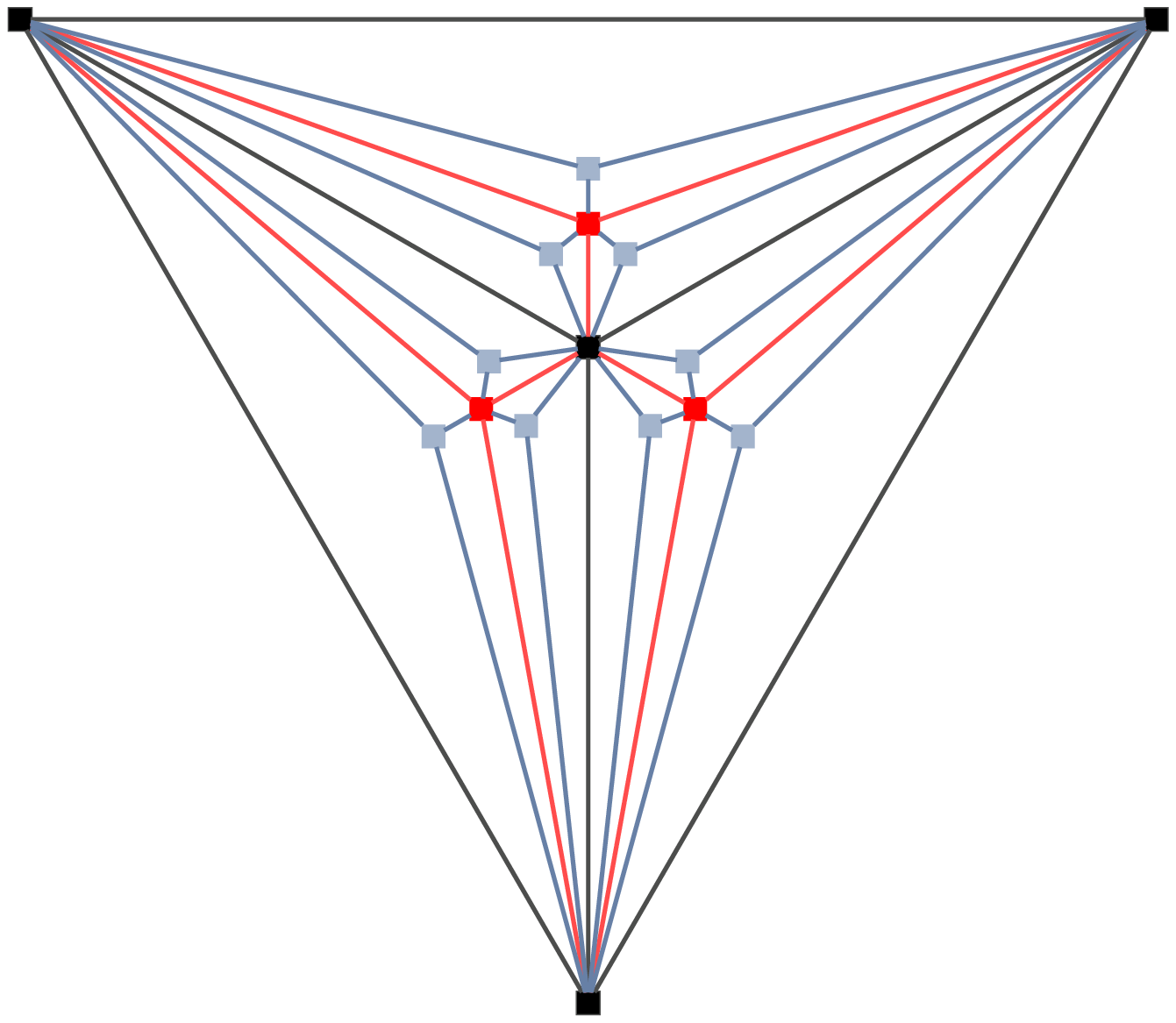}} &
{\includegraphics[width=0.45\columnwidth]{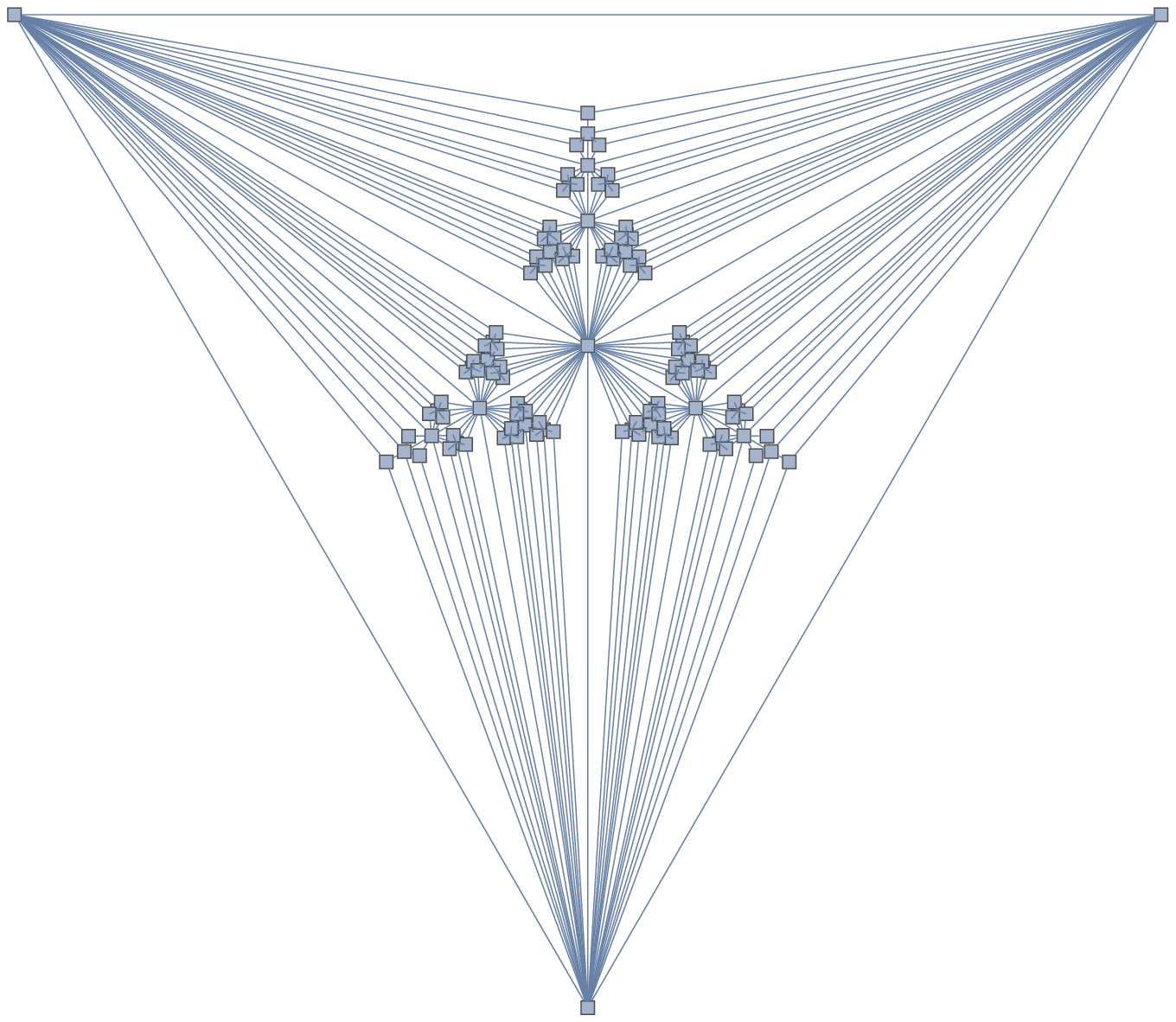}}
\end{array}$
\caption{(Color online) Left panel: the first 3 generations of the Apollonian graph. The nodes and links added to construct the 1st, 2nd  and 3rd generations are shown. Right panel: the 5th generation Apollonian network.}
\label{apollonian}
\end{figure}
\subsection{Apollonian networks}

\begin{figure}[!h]
\center
\includegraphics[width=1.0\columnwidth ]{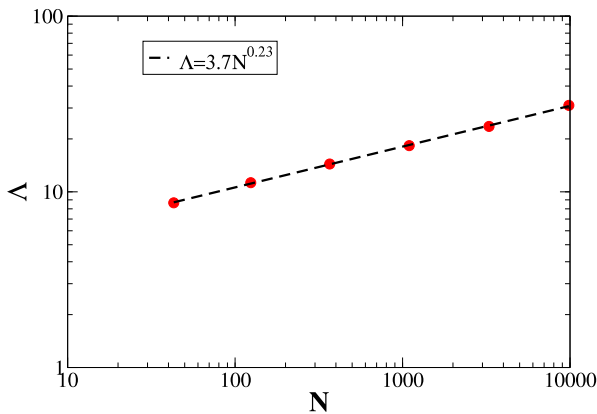}
\caption{Scaling of the maximal eigenvalue $\Lambda$ of the Apollonian network as a function of the size $N$ of the network.}
\label{apollonian_eigen}
\end{figure}

We consider here Apollonian networks \cite{Apollonian1} which are constructed through a 2D Apollonian packing model 
 in which the space between three tangent circles placed on the vertices of an equilateral triangle is filled by a maximal circle. The space-filling procedure is repeated for every space bounded by three of the previously drawn tangent circles. The corresponding Apollonian network is constructed by connecting the centers of all the touching circles [Figure \ref{apollonian} (left)].
The resulting network is scale-free with power-law degree distribution $p(k)={\cal N}k^{-\lambda}$ and $\lambda=1+\ln(3)/\ln(2)\simeq 2.585$. Also these networks are known to have diverging maximal eigenvalue $\Lambda$ of their adjacency matrix \cite{Apollonian2}. In Figure $\ref{apollonian_eigen}$ we plot the maximal eigenvalue $\Lambda$ of the apollonian network as a function of the network size $N$. We can fit the numerical results with the function
\bea
\Lambda(N)\propto N^{0.23}.
\eea
Therefore the hopping coefficient $\kappa$ [that needs to scale according Eq. ($\ref{scaling}$)] scales for large $N $ as
\bea
\kappa(N)\propto N^{-0.23}.
\eea

\subsection{Small-world networks}
\begin{figure}
\center
\includegraphics[width=1.0\columnwidth ]{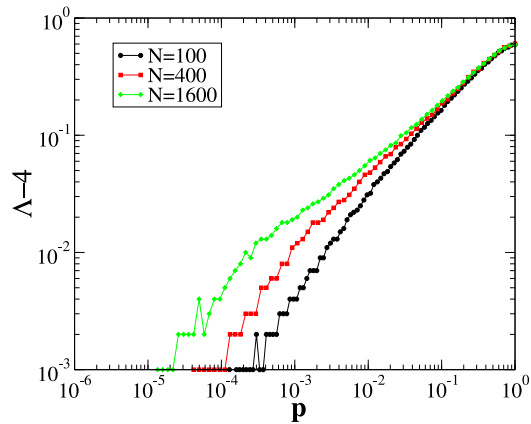}
\caption{The maximal eigenvalue of the Small-World network as a function of $p$ for different network sizes $N$. In the limit $p\to 0$ the small world model is a regular chain with average connectivity $z=4$, therefore in this limit $\Lambda=4$. The data are averaged over 100 network realizations.}
\label{small_world}
\end{figure}

Small-world networks structures characterize a system that has at the same time a small diameter, like random networks, but {also} has a high clustering coefficient, {similarly to} regular networks \cite{SW}. 
In particular, we can follow the construction proposed in \cite{SW}: we start from a regular chain in which each node is linked to the nearest neighbours and to the next-nearest neighbours; then every link is rewired with probability $p$ to another random node of the network.
For $p=0$ the small-world network is a regular lattice in one dimension; for $p=1$ the small-world network becomes one instance of a random graph network; finally, for every intermediate value of $p$ we observe the small-world network with small average diameter and a high clustering coefficient.
The maximal eigenvalue of this network, for $p=1$ will increases with $N$, as in the random graph case [Eq.(\ref{ER})], while for the case $p=0$ it will be independent on $N$, as in regular networks.
In Figure $\ref{small_world}$ we see how the maximal eigenvalue of the network changes with $N$ for intermediate values of the probability $p$ when the network is small-world.

\section{ Conclusions}\label{sec:conclusion}
In conclusion we have studied the Jaynes-Cummings-Hubbard (JCH) model on complex quantum networks.
We have shown that the phase diagram derived in the mean-field approximation, depends crucially on the maximal eigenvalue $\Lambda$ of the hopping matrix.
In particular the phase diagram depends on the product $\kappa \Lambda$. This implies that in order to have a well defined phase diagram in the large network limit, the hopping coefficient $\kappa$ should scales proportionally to $1/\Lambda$.
The eigenvalue $\Lambda$ is equal to the connectivity of the network, for regular networks and lattices, but for complex random networks it generally increases with the network size.
In this paper we have listed for a large class of networks the scaling of the maximal eigenvalue $\Lambda$ with the network size $N$.
For complex networks that have a diverging $\Lambda$, the hopping coefficient should be a decreasing function of $N$ in order to observe {the} phase transition  {from the Mott-like regime to the superfluid phase. This result implies the possibility of observing quantum critical behaviours in arrays whose cavities are weakly coupled, assuming the effective realization of the proper complex quantum network topology.}

\appendix
\section{Mean field solution of the JCH model}
The Jaynes-Cummings Hamiltonian (we set $\hbar=1$) 
\beq
H^{JC}=\epsilon \sigma^+\sigma^- + \omega a^\dagger a + \beta(\sigma^+a+\sigma^-a^\dagger)
\eneq
is obtained in the rotating wave approximation and in the limit $\beta \ll \epsilon, \omega
$. The total number of excitations is a conserved quantity, and it is given by the sum of electromagnetic and atomic excitations $N=a^\dagger a +\sigma^+\sigma^-$. The interacting part of the Hamiltonian connects $n$-sectors which differs only by one photon excitation 
\beq
\ket{n-1}\upket \longleftrightarrow  \ket{n}\downket
\eneq
The JC Hamiltonian can then be block-diagonalized in different sectors, each labelled by $n$, and each sector spanned by $\{\ket{n-1}\upket, \ket{n}\downket \}$. Choosing this set as the basis for the $n-$th sector, Eq.(\ref{eigve}) and Eq.(\ref{eigva}) in the main text provide the expressions for the eigenvectors and eigenvalues of the JC Hamiltonian. 

Considering the situation of a network of cavities, whose coupling is effectively described by an hopping term, we have the Jaynes-Cummings-Hubbard Hamiltonian described in Eq.(\ref{eq:JCH}). As explained in the main text, the mean-field treatment of the hopping term allows us to approximate the JCH Hamiltonian as follows
\beqynn
H^{MF}&=&\sum_i \epsilon \sigma_i^+\sigma_i^- + \omega a_i^\dagger a_i + \beta(\sigma_i^+a_i + \sigma_i^-a_i^\dagger) \\
&&-\kappa \sum_{i,j}\tau_{ij} (a^{\dag}_i+a_i)\psi_j+\kappa \sum_{i,j}\tau_{ij}\psi_i\psi_j.
\eneqynn
Considering the hopping term as a perturbation to the atomic limit, the order parameter of the model is provided by $\psi_i\equiv \langle a_i \rangle_{gs}$, where the expectation value is calculated with respect to the ground-state of the Jaynes-Cummings-Hubbard model to first order in perturbation theory. Note that the order parameter can be assumed real due to the gauge symmetries of the Hamiltonian \cite{LeHur}. The self-consistent equation for the order parameter can then be written as 
\begin{equation}
\psi_i \equiv \langle n_{1}|a_i|n_{1}\rangle,
\label{Eq:selfc}
\end{equation}
where $|n_{1}\rangle \equiv |n,-\rangle^0 + |n\rangle^1 $ is the approximation of the 
ground-state to first-order in the perturbation, while $|n,-\rangle^0$ is the ground-state 
of the unperturbed Hamiltonian (see Eq.\ref{eigve} in the main text), and 
$$
|n\rangle^1 = \sum_{k,\alpha=\pm} \frac{^0\langle k,\alpha|H^{MF}_{hop}|n,-\rangle^0}{E^\mu_{n,-}-E^\mu_{k,\alpha}}|k,\alpha\rangle^0.
$$ 
From Eq. (\ref{Eq:selfc}) we have 
\beqynn
\langle n_{1}|a_i|n_{1}\rangle &=& ^0\langle n,-|a_i|n,-\rangle^0+\,^1\langle n|a_i|n\rangle^1\\
&&+\,^0\langle n,-|a_i|n\rangle^1+\,^1\langle n|a_i|n,-\rangle^0.
\eneqynn
Keeping only non-zero terms to first order in $\kappa$ we are left only with the last two terms in the above equation. 
First we explicitly calculate 
$$
^0\langle n,-|a_i|n\rangle^1
=\sum_{k,\alpha=\pm} \frac{^0\langle k,\alpha|H^{MF}_{hop}|n,-\rangle^0}{E^\mu_{n,-}-E^\mu_{k,\alpha}}
\,^0\langle n,-|a_i|k,\alpha\rangle^0.
$$
It is easy to check that the only non-zero terms in the sum are given by $k=(n+1)$, and $\alpha=+,-$. It follows that the non-zero contribution to the expectation value of $H^{MF}_{hop}$ is provided only by $-\kappa\sum_j \tau_{ij} \hat{a}_i^\dagger \psi_j$.
From the explicit form for the ground-states of the unperturbed Hamiltonian (see Eq.\ref{eigve} in the main text) we obtain
\beqynn
^0\langle n,-|a_i|(n+1),-\rangle^0&=&\sqrt{n+1}\cos{\theta_n}\cos{\theta_{n+1}}\nonumber \\
&&+\sqrt{n}\sin{\theta_n}\sin{\theta_{n+1}}
\eneqynn
\beqynn
^0\langle n,-|a_i|(n+1),+\rangle^0&=&\sqrt{n+1}\cos{\theta_n}\sin{\theta_{n+1}}\nonumber\\
&&-\sqrt{n}\sin{\theta_n}\cos{\theta_{n+1}}.
\eneqynn
We can proceed similarly for the calculation of $^1\langle n|a_i|n,-\rangle^0$, obtaining
\beqynn
^0\langle (n-1),-|a_i|n,-\rangle^0&=&\sqrt{n}\cos{\theta_n}\cos{\theta_{n-1}}\nonumber\\
&&+\sqrt{n-1}\sin{\theta_n}\sin{\theta_{n-1}}
\eneqynn
\beqynn
^0\langle (n-1),+|a_i|n,-\rangle^0&=&\sqrt{n}\cos{\theta_n}\sin{\theta_{n-1}}\nonumber \\
&&-\sqrt{n-1}\sin{\theta_n}\cos{\theta_{n-1}}.
\eneqynn
Now let us define
\beqynn
R_n &\equiv &- \sum_{\alpha=\pm}\left[\frac{|^0\langle n,-|a_i|(n+1),\alpha\rangle^0|^2}{E^\mu_{n,-}-E^\mu_{(n+1),\alpha}}\right.\\
&&+\left.\frac{|^0\langle (n-1),\alpha|a_i|n,-\rangle^0|^2}{E^\mu_{n,-}-E^\mu_{n-1,\alpha}}\right].
\eneqynn

Putting everything together we have the following self-consistent equation to first order in perturbation theory (for $n>0$)
$$
\psi_i=\kappa R_n\sum_j \tau_{ij}\psi_j,
$$

The case $n=0$ can be calculated in the same way and one has
$$
\psi_i=\kappa R_0\sum_j \tau_{ij}\psi_j,
$$
where 
$$
R_0=-\sum_{\alpha=\pm}\frac{|\langle 0,\downarrow|a_i|1,\alpha \rangle|^2}{E^\mu_{0,\downarrow}-E^\mu_{1,\alpha}},
$$
with
$
\langle 0,\downarrow|a_i|1,- \rangle=\cos{\theta_1}
$, and 
$
\langle 0,\downarrow|a_i|1,+ \rangle=\sin{\theta_1}.
$

\end{document}